**Resurrecting Socrates in the Age of AI: A Study Protocol for Evaluating a Socratic Tutor to Support Research Question Development in Higher Education**

Ben Degen

Faculty of Human Sciences, University of Kassel

**Author note**

Ben Degen 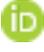 https://orcid.org/0000-0002-4330-0893

## Abstract

Formulating research questions is a foundational yet challenging academic skill, one that generative AI systems often oversimplify by offering instant answers at the expense of student reflection. This protocol lays out a study grounded in constructivist learning theory to evaluate a novel AI-based Socratic Tutor, designed to foster cognitive engagement and scaffold research question development in higher education. Anchored in dialogic pedagogy, the tutor engages students through iterative, reflective questioning, aiming to promote System 2 thinking and counteract overreliance on AI-generated outputs. In a quasi-experimental design, approximately 80 German pre-service biology teacher students will be randomly assigned to one of two groups: an AI Socratic Tutor condition and an uninstructed chatbot control. Across multiple cycles, students are expected to formulate research questions based on background texts, with quality assessed through double-blind expert review. The study also examines transfer of skills to novel phenomena and captures student perceptions through mixed-methods analysis, including surveys, interviews and reflective journals. This study aims to advance the understanding of how generative AI can be pedagogically aligned to support, not replace, human cognition and offers design principles for human-AI collaboration in education.

## Keywords

## 1. **Introduction**

Well-formulated research questions are crucial for academic research, shaping the trajectory and scope of scientific inquiry, helping to define hypotheses and guiding experimental design (Covvey et al., 2024; Lipowski, 2008). However, students at higher education institutions (HEIs) often struggle to develop clear and focused research questions, particularly those who are novice researchers (Booth et al., 2016). (Under-)graduates frequently encounter challenges in crafting these questions due to limited experience in synthesising broad knowledge into focused queries and often struggle with the translation of complex ideas into researchable terms.

Acknowledging this, remedies were sought after by lecturers of HEIs all over the world. Efforts have been made by systematically linking research and teaching, e.g. by creating inquiry-based learning modules requiring students to use practices of professional researchers (Böttcher & Thiel, 2018; Lehmann & Mieg, 2018; Pedaste et al., 2015). Others proposed various pedagogical approaches that could be utilized, including structured guidance through worksheets, scaffolding techniques, and explicit instruction on research question development (Byrd & Camba, 2020; Kanter & Byrd, 2020; Zheng & Byrd, 2020).

Due to the resource-intensive support required, Artificial Intelligence (AI), i.e. "computing systems that are able to engage in human-like processes such as learning, adapting, synthesizing, self-correction, and use of data for complex processing tasks" (Popenici & Kerr, 2017, p. 2) has emerged as a promising tool for enhancing higher education (Zawacki-Richter et al., 2019). However, in the author's view many AI-based solutions fall into a trap formulated by Popenici and Kerr in 2017 on using AI in higher education. The tendency to (…) "look at technological progress as a solution or replacement for sound pedagogical solution or good teaching." (p.3). Furthermore, more recent research also shows, that uncontrolled usage of AI technologies may increase dependence on generative AI and potentially cause "metacognitive laziness" (Fan et al., 2025). This paper lays out an experiment integrating sound pedagogics into AI that aims to bridge this trap and counter metacognitive laziness. The here proposed Socratic dialogue-based approach aims to stimulate deeper learning through a series of probing questions.

## 2.  Background

### 2.1. Scaffolding research question generation at higher education institutions

For students in HEIs, research question development is a crucial skill that often requires structured guidance and support (Booth et al., 2016; Covvey et al., 2024; Lipowski, 2008). Although there are several frameworks with varying details to describe the research question development process, two basic steps for which support has been developed can be distilled from the broader literature across research domains.

The initial step in research question development involves identifying an interesting topic (Booth et al., 2016; Kanter & Byrd, 2020; Zheng & Byrd, 2020). This often requires students to move from broad areas of interest to more focused topics that can be effectively explored within the scope of a research project. To support this transition, methods such as activity worksheets have been developed. For instance, the worksheet method described by Byrd and Camba (2020) employs a structured process of identifying topics by guiding students to articulate their existing knowledge about potential topics, identifying gaps in their understanding and areas requiring further investigation as well as identifying the significance of a topic by employing rankings. The helpfulness of this incremental scaffolding approach has been positively evaluated by graduate students, especially those in their 1st semester.

Once a topic has been identified, the next challenge is to transform it into a well-defined research question. Some suggest practical heuristics, such as asking "why" five times to get to "the heart of the issue" (Lipowski, 2008, p. 4) or to use the four s's: size, scope, scalability and sustainability (ibid). Other researchers highlight the importance of posing questions that are not only significant, addressing an important gap in the existing knowledge, but also specific, with a clearly defined scope and focus, and answerable, i.e. capable of being investigated through appropriate research methods. Additionally, students should consider the audience for their research and the broader context of their research community. This entails understanding disciplinary expectations, such as the types of questions valued within their field of study; research conventions, including the accepted formats for framing and presenting inquiries; and, crucially, prior research, encompassing the existing body of knowledge and the ongoing scholarly conversations within their research community.

Understandably, this set of expectations can be daunting for students and is seldom met. Traditionally HEIs rely on academic staff to provide support and alleviate some of the students' concerns despite the high (opportunity) cost. Recognising the challenges students face in framing meaningful and well-defined research questions, as well as the resource intensiveness of the support, educators and institutions have explored AI in recent years to offer solutions.

## 2.2. AI-supported Research Question Development

Utilizing intelligent tutoring systems (Aleven et al., 2006; Dermeval et al., 2018; Pardos et al., 2023; Woolf, 2009), and adaptive learning platforms (Cavalcanti et al., 2021), AI is used in various scenarios to personalize learning experiences (Du & Daniel, 2024; Holmes et al., 2019), and to provide feedback to students and teachers alike (Demszky & Liu, 2023; Meyer et al., 2024; Morris et al., 2024).

With regard to AI's role in supporting research question development, current approaches to AI assistance can broadly be categorised into two types: AI-Centric Generation and Human-AI Collaboration. Here it is argued that the former approach is not fit for users' educational progress. Indeed, as will be shown, many of the currently available AI tools for research question development are not explicitly designed for educational purposes, even those AI tools that are intended to act as support.

Research published by Lu and colleagues (2024) for example falls into AI-Centric-Generation category. Their paper, called *The AI Scientist: Towards Fully Automated Open-Ended Scientific Discovery* lays out an AI-based software that mimics researchers' process: "The AI Scientist seamlessly performs ideation, a literature search, experiment planning, experiment iterations, manuscript writing, and peer reviewing to produce insightful papers." (p. 2). Another example is the work on idea generation by Si et al. (2024). The three researchers from Stanford University explored whether researchers or an LLM ideation agent would produce more novel research ideas. They recruited over 100 researchers to write novel ideas and carried out blind reviews of both LLM and human ideas. The findings were very startling: "(…) we conclude that AI ideas generated by our ideation agent are judged as more novel than human expert generated ideas, consistently across all three different statistical tests." (p.11). Although one should mention, that the LLM ideas were considered more novel and slightly less feasible than those of human experts.

While both examples demonstrate astonishing accomplishments, it is virtually non-useful or even detrimental for educational purposes, as it defies the purpose of education to foster the skills of students and promote deeper engagement with the questions at hand as it offloads critical thinking to the AI.

The challenge for researchers in education therefore is to use what we know about learning and teaching to create AI applications for educational design purposes. Here it is argued that this can only be accomplished via Human-AI-Collaboration.

One example for this is the deployment of recommender systems. Utilizing an LLM-based agent system called CoQuest Liu and colleagues (2023) sought to support research question development by automatically generating new research questions for users to act on.

Despite CoQuest having students in mind, there are still problems remaining that limit the impact regarding fostering the skills of students. First, recommender systems might lead to overreliance, as their basis is reliance on the recommendations. This overreliance appears to be hard to overcome, as a study by Buçinca et al. (2021) demonstrated. Intended to reduce overreliance by adding explanations to recommendations, Buçinca and her colleagues had no substantial success: "(…) when the AI suggests incorrect or suboptimal solutions, people still on average make poorer final decisions than they would have without AI's assistance" (Buçinca et al., 2021, p188:2).

Second, students might use rather quick, intuitive thinking and heuristics to quickly analyse the quality of the presented recommendations rather than deliberate thinking. This reliance on intuitive, fast, and automatic judgments aligns with the work of Kahneman, Frederick and Tversky on the concept of System 1 thinking, which operates effortlessly but is prone to biases and errors (Kahneman, 2011; Kahneman & Frederick, 2005; Tversky & Kahneman, 1974). This assumption is further supported by current research on the use of ChatGPT, highlight a usage pattern that favours easy tasks, such as summarisation, rather than fostering thoughtful analysis and reflection (Ravšelj et al., 2025). In contrast, deliberate and effortful System 2 thinking would enable a more critical evaluation of the AI's outputs. However, fostering System 2 thinking in students when interacting with AI tools requires not only recommendations but a design that encourage reflective engagement rather than impulsive acceptance of recommendations.

The challenge lies in creating AI systems that mitigate the cognitive biases inherent to System 1 while scaffolding the deliberate, critical reasoning processes associated with System 2. The next section lays out one possible solution to this challenge which builds upon a traditional pedagogical method that inherently foster deliberate, reflective, and analytical reasoning: Socratic questioning.

### 2.3. The Socratic method

Dialogues, in the form of vigorous debates, question and answer, criticism and opinion have ever since been tools of educators to promote skills. Particularly useful has been the instructional function of dialogue, i.e. "an intentional process in which a teacher "leads" a student through questioning and guidance, to formulating certain answers or understandings" (Burbules & Bruce, 2001, p. 1122). This is also often referred to as *Socratic Questioning*.

Socratic questioning has a long history in education, dating back to ancient Greece and its namesake Socrates (Heckmann & Krohn, 2018) and is one of many types of dialogic methods with kinship to Socrates' method (Knezic et al., 2010). Despite well-documented debates on the origins and actual usage by Socrates (Burbules & Bruce, 2001), discord due to the ambiguous usage of the term (Carey & Mullan, 2004), and differences in definition (Lee et al., 2014) common elements can be found within the literature.

First, a guiding facilitator is needed, whose task it is to stimulate and steer the conversation. Second, The questioning process traditionally followed a cross-examination format, but evolved into a cooperative exploration between the questioner and the individual being questioned (Overholser, 1993). Third, the focus lies on guiding students towards a deeper understanding of a given topic rather than simply providing them with direct answers, which might explain its effectiveness in promoting higher-order thinking and developing competencies (Fahrner & Wolf, 2020; Katsara & De Witte, 2019; Knezic et al., 2010; Yang et al., 2005). Fourth, a series of questions is aimed at guiding individuals to discover knowledge through logical reasoning and/or self-reflection dissociating socratic *questioning* from the socratic *dialogue*, which aims to stimulate a group.

The types of questions used by the facilitator might differ largely and depend on the goals to be achieved. Paul (1990, p. 276-278) for example lists six categories and examples of generic questions for each category in his *Taxonomy of Socratic Questions*, as listed on the following page. Interested readers might also look at the revised, but in the authors' view more generic, taxonomy published by Paul & Elder in 2007.

Table 1. Taxonomy of Socratic Questions including selected examples of questions.

| Category | Selected examples |
| --- | --- |
| Questions of clarification | What do you mean by __________? Could you give me an example? Could you explain that further? Could you put that another way? How does this relate to our discussion (problem, issue)? |
| Questions that probe assumptions | What are you assuming? What could we assume instead? You seem to be assuming. How would you justify taking this for granted? Is it always the case? |
| Questions that probe reason and evidence | What would be an example? Are these reasons adequate? Do you have any evidence for that? How does that apply to this case? But is that good evidence to believe that? |
| Questions about viewpoints or perspectives | You seem to be approaching this issue from perspective. Why have you chosen this rather than that perspective? How would other groups/types of people respond? Why? What would influence them? What would someone who disagrees say? |
| Questions that probe implications or consequences | What are you implying by that? What effect would that have? Would that necessarily happen or only probably happen? What is an alternative? |
| Questions about the question | How can we find out? What does this question assume? Why is this question important? Can we break this question down at all? Is the question clear? Do we understand it? Is this question easy or hard to answer? Why? |

*Note.* By Paul (1990, p. 276-278)

While those questions were traditionally employed in interpersonal face-to-face interactions, this study tries to adapt the Socratic questioning method for AI-based interactions.

## 2.4. Constructivist theoretical assumptions

The AI-based Socratic questioning draws on fundamental constructivist theoretical assumptions. Vygotsky's concept of the *More Knowledgeable Other* suggests that effective learning occurs when guidance is provided by a more experienced entity (Vygotsky, 1980). Be it a human instructor or, in this instance, an AI system. Similarly, Bruner's Spiral Curriculum (1966) suggests that learning is most effective when learners construct their knowledge through exploration and the subject at hand is revisited at increasing levels of complexity, allowing students to deepen their understanding over time. An AI Socratic Tutor can align with these principles by engaging students in iterative dialogues that progressively refine their research questions, encouraging them to revisit and expand upon their initial ideas.

Hence, a fundamental assumption of this work is, that an AI chatbot instructed to use Socratic questioning can foster deeper engagement with the subject matter and encourage students to construct knowledge collectively through reasoned discourse, presenting an exciting avenue for enhancing learning and fostering deeper engagement. An experiment is warranted to test this assumption, with the following research questions delivering guidance for the experiment:

## 3. Research Questions

- RQ1: To what extent does the use of a Socratic AI Tutor improve the quality of research questions formulated by students compared to an uninstructed AI chatbot?

- RQ2: How do students perceive the educational value of the Socratic AI Tutor compared to an uninstructed AI chatbot?

- RQ3: To what extent does the Socratic AI Tutor promote the transfer of research question development skills to novel phenomena compared to an uninstructed AI chatbot?

## 4. Methodology and materials

### 4.1. Research approach

The methodology employed for this study follows a mixed-method, quasi-experimental design to ensure a comprehensive evaluation of the effects of the socratic instruction. Two courses of students enrolled in a Biology didactics course at an university in Germany, totalling approximately 80 graduate teacher education students, will be assigned to one of the following groups:

- **Group 1**: Students using the Socratic AI Tutor.

- **Group 2**: Students using an uninstructed AI chatbot.

Students of both groups will use the chatbots in an iterative process, repeating the intervention spanning the semester, as laid out in the next section.

### 4.2. Procedure

Initially, the students of both groups will generate codes for the login to the digital survey environment used to carry out the experiment. These codes provide an anonymised unique identification across the whole semester. Furthermore, a pre-test on their research question development competency will be conducted. The test will be based on the sub-dimension *Formulating questions* of the test instrument developed in the longitudinal multi-cohort study Ko-WADiS (Mathesius, 2014). Each following iterative cycle will be comprised of two parts. Part 1 is the participation in the survey environment which encompasses the following stages:

1. provision of background texts
2. prompting the students to develop an initial research question
3. forwarding to the refinement of the research question supported by either the Socratic AI Tutor or the uninstructed AI chatbot
4. testing via a transfer task
5. collection of survey data

Part 2 utilises learning journals gain insights on research question development and the use of the AI support via meta-reflection

The phenomena for research question development are pre-selected and a background text is provided to all participants to focus their efforts and increase comparability of the results. Furthermore, the phenomena will gradually increase in their degree of abstraction towards broader topics, to accommodate for anticipated initial problems in creating research questions. The provisioned background texts are purposefully designed to start narrow but do allow the development of multiple areas of interest at a later stage. Similarly, the transfer task will be designed similarly with background texts covering a different phenomenon. An exemplary background text for a later stage and exemplary research questions can be found in Annex A.

The first stage of part 1, i.e. the provision of background texts is designed in a way, that students are only allowed to continue after two minutes to ensure that the text has been read and prevent prematurely skipping. These background texts will also be available during the following stage of initial research question.

### 4.3. Hypotheses

Based on the conducted literature view and the assumption stated earlier, the following hypotheses have been deducted:

- H1: Research questions developed with the support of the Socratic AI Tutor will be significantly better rated than initially submitted research questions.

  o H1.1: The difference between the ratings of initially developed research questions and the improved questions in the group with the support of the Socratic AI Tutor will decrease over time.

Based on the work of Si et al. (2024), which found that AI generated research questions were often judged better than questions developed by researchers, as mentioned earlier, H2 has been added:

- H2: Research questions developed by students using the Socratic AI Tutor will be rated lower than research questions developed by students in the uninstructed chatbot group.

- H3: Students will report higher levels of perceived cognitive engagement and reflection when using the Socratic AI Tutor compared to uninstructed chatbot group.

- H4: The Socratic AI group will demonstrate higher performance in transfer tasks requiring the formulation of research questions in a different observation/topic compared to the uninstructed chatbot group.

### 4.4. The AI-Based Socratic Tutor

Initial attempts to fine-tune an open instruction model (Llama-3.1-8B) on a semi-synthetic dataset using monolithic odds ratio preference optimization (ORPO) (Hong et al., 2024). For interested readers the code used for training as well as the database and the model can be accessed under https://colab.research.google.com/drive/10PnLhkPpDI_8CLayI3-

75h9MMznc11cL?usp=sharing and https://huggingface.co/QuestforAIEd], respectively.

These attempts led to disappointing results, most definitely due to the low number of

examples within the dataset. Due to the lack of a viable dataset, the ChatGPT Assistant by

OpenAI based on the gpt-4o model (03/2025) was utilized.

Testing different system instructions led to the inclusion of several aspects of a

Socratic dialogue and the PICOT-framework (Riva et al., 2012). The PICOT framework,

originally stemming from the field of medicine, guides formulating precise research questions, particularly in evaluating interventions. It comprises five elements: Population (P),

specifying the target group for the study; Intervention (I), detailing the treatment under

investigation; Comparison (C), identifying the control or alternative intervention; Outcome

(O), describing the measurable effects of the intervention; and Time (T), indicating the duration for assessing outcomes. Ultimately, the assistant was embedded into the digital survey environment via API.

The temperature, i.e. the setting that allows to control for randomness when picking words during text creation on a scale of 0 to 1, with low values making the text more predictable but less creative, was set to 0.10 for answers to be logical and consistent. In addition, the Top-P-value, i.e. the number of words possibly to be considered by the model, was set to 0.50 to allow for answers in diverse contexts but in a consistent manner. These settings were chosen based on the work by Amin & Schuller (2024) on sensitivity analysis, who found that "(…) conservative predictions with lower T ≤ 0.3 values or top- p ≤ 0.7 yield better and stable performance. Increasing T or Top-P beyond that generally worsened the performances" (p. 6). The user interface (see Fig. 1) was designed to minimize distraction and with commonly known designs of AI chats in mind to increase usability.

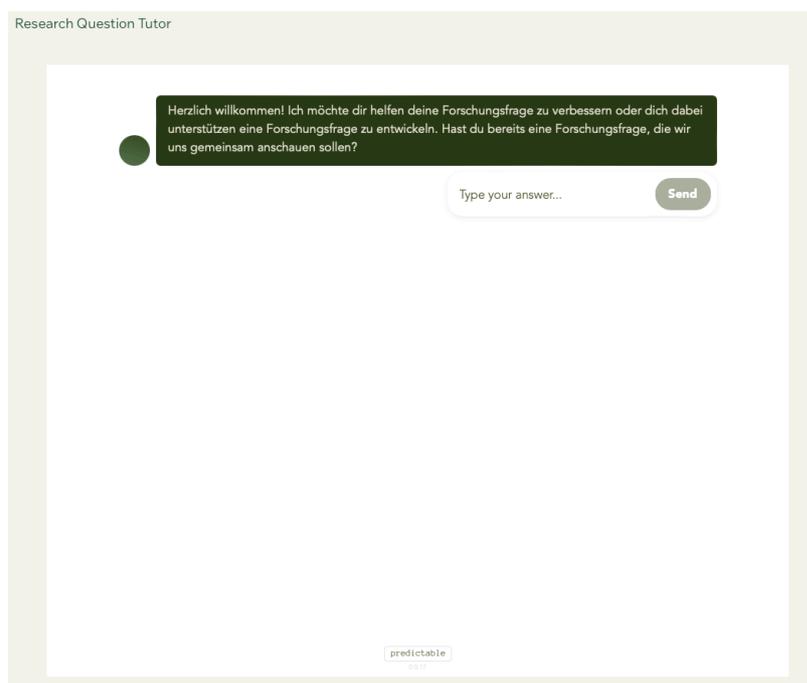

Fig. 1. User interface of the AI-based Socratic Tutor.

## 4.5. Data Collection

A Likert-scale questionnaire will be employed to gather information on students' general perceptions and usage of AI chatbots, as well as their specific experiences with the Socratic Tutor. To ensure methodological robustness and contextual relevance, the questionnaire was developed by integrating elements from the Unified Theory of Acceptance and Use of Technology (UTAUT; Venkatesh, 2022) and the validated survey instrument by Stöhr et al. (2024).

The scale was designed to balance predictive validity with practical usability, minimising participant fatigue while maximising informational yield and is expected to allow for a nuanced assessment of both students' acceptance of the Socratic AI Tutor and its educational effectiveness. Demographic information (e.g., gender, academic level) will also be collected to enable subgroup analysis. In addition to the quantitative measures, the survey includes open-ended questions to capture students' subjective impressions.

User queries and assistant responses are collected using the ChatGPT Assistant platform, cleaned, and stored in a tabular format to prepare for further data analysis. In addition, the students learning journal entries as well as the output of additionally implemented guided interviews will be collected to receive an in-depth understanding of the intervention outcomes. A data management plan was developed to be made public via the Verbund Forschungsdaten Bildung (German Network of Educational Research Data).

### 4.6. Data Analysis

The quality of the initially submitted research questions as well as the final submitted research questions of each iteration will be assessed utilising double-blind review by faculty of the university and a 0-15 point scale, which is commonly utilised in grading course work in Germany, whereas 15 is the highest achievable outcome. Inter-rater reliability will be calculated using Fleiss' Kappa (Fleiss & Cohen, 1973).

Quantitative data, such as the quality of research questions, will be analysed using ANCOVA to identify significant differences regarding the pre- and post-intervention-quality of the research questions between the two groups. Descriptive and inferential statistics will be applied to the survey data to assess students' perceptions of the tool's usefulness, with responses summarized and compared across groups. A thematic analysis to uncover recurring patterns in the qualitative data is planned, and it is expected to gain more nuanced perspectives on students' interactions with the Socratic Tutor.

## 5.  ETHICS:

This study adheres to rigorous ethical standards to ensure the protection of participants. Informed consent was obtained from all participants prior to their involvement, clearly outlining the purpose of the study, the voluntary nature of participation, and the right to withdraw at any time without penalty (Israel & Hay, 2006). Anonymity was ensured by assigning unique identification codes to participants. Care was taken to minimise potential risks, such as stress or discomfort during the intervention, by providing clear instructions and an opportunity to ask questions throughout the process. We acknowledge the fact, that intensive use of new technology can lead to intervention-generated inequalities, benefitting socioeconomically advantaged groups (Veinot et al., 2018). Hence, it is planned to grant access to the Socratic Tutor to all students at the HEI in case it proves to be successful in supporting research question development. Interviews are conducted on a strictly voluntary basis. Furthermore, the study is conducted in a course that is not taught by the author to prevent dependency effects.

## 6. Limitations

Despite best efforts, several limitations warrant consideration. The use of a proprietary large language model (ChatGPT-4o) presents a known constraint: the model's internal updates and decision-making processes are not fully transparent to users. However, this choice was made deliberately, as initial attempts at fine-tuning an open-source model proved insufficient for achieving pedagogically coherent Socratic dialogues. The use of commercial tools reflects a broader challenge in aligning generative AI systems with educational goals and research transparency. Additionally, the study's focus on pre-service biology teacher education within a single institutional setting may limit generalisability to other domains or cultural contexts. Finally, individual learner differences such as prior subject knowledge, comfort with AI, and self-regulation may influence both engagement and outcomes but may not be sufficiently captured by the methods deployed.

## 7. Conclusion

This study protocol proposes a study for a novel approach to leveraging generative AI for educational purposes by integrating the principles of Socratic questioning into an AI-powered chatbot. Recognising the limitations of AI-centric generation models, which often prioritise efficiency at the expense of critical engagement, an experimental design was outlined. It seeks to promote reflective inquiry and support the development of research question formulation skills among pre-service teacher education students.

Grounded in constructivist learning theory, the Socratic AI Tutor is conceptualised not as a provider of answers, but as a facilitator of thinking. Its design seeks to counteract tendencies toward metacognitive laziness and to scaffold System 2 reasoning through structured questioning. The study follows a mixed-method research design to examine not only measurable outcomes such as the quality of student-generated research questions and transfer of skills, but also the nuanced experiences and perceptions students hold about interacting with AI in a reflective manner.

While the empirical phase of the study is yet to be conducted, the theoretical foundations and methodological considerations suggest a promising step toward more thoughtful and pedagogically aligned uses of AI in higher education. In contrast to many current AI implementations, this work emphasises the *process* of learning over the *product*, treating the learner not as a passive recipient of machine intelligence, but as an active participant.

Ideally, this protocol may serve as a foundation for future research on the design and deployment of AI tools that enhance, not replace, human cognition. The aim is to advance a more pedagogically sound, and intellectually engaging vision for AI in education.

## 8.  Future avenues for investigation

Regardless of the results, further work will be necessary, and several avenues might warrant further exploration. First, the design of the tutor itself might merit exploration. As delaying the presentation of AI recommendation has been shown to enhance reflective (Liu et al., 2023; Park et al., 2019), it is worth investigating whether the immediacy of AI feedback might inadvertently prioritise perceived helpfulness over thoughtful engagement. Strategies to encourage deeper cognitive processes, such as intentional delays, aligning with the dual-process model of System 1 and System 2 (Kahneman, 2011), might help in achieving educational outcomes. Furthermore, the time available to develop research questions might be expanded to allow for more thoughtful engagement, too.

Secondly, planning research on the domain-specificity of the Socratic tutor, i.e. testing whether it's helpful in other domains such as math or chemistry, by employing generalizability studies. This also leads to the second avenue, the transferability of the skillset: does the Socratic tutor foster research question development skills in such a way that students can apply them across disciplines and real-world problem-solving contexts? Investigating whether the inquiry-driven mindset promoted by the tutor transfers to non-academic settings would offer valuable insights into its broader applicability.

Further avenues for research include exploring the longitudinal effects of Socratic tutor use. While short-term studies may reveal improvements in research question formulation, examining whether these skills persist and evolve over time could illuminate the tutor's lasting impact. Additionally, studying the role of individual differences, such as cognitive style, self-regulation, and prior knowledge, could help tailor the Socratic tutor to diverse learner profiles, ensuring its effectiveness across varying educational contexts.

ANNEX A

Example of a version of a provisioned background text for the late stage

---

**Fledermäuse – Meister der Nacht**

Fledermäuse sind die einzigen Säugetiere, die aktiv fliegen können. Weltweit existieren über 1.400 Arten, die eine zentrale Rolle in verschiedenen Ökosystemen spielen. Sie ernähren sich je nach Art von Insekten, Früchten, Nektar oder sogar Blut. Insektenfressende Fledermäuse sind wertvolle Schädlingsbekämpfer, da sie große Mengen an Insekten konsumieren. Ihr Orientierungssystem, die Echoortung, ermöglicht es ihnen, sich auch in völliger Dunkelheit sicher zu bewegen. Dabei stoßen sie hochfrequente Schallwellen aus, die von Objekten reflektiert werden und so ein akustisches Bild ihrer Umgebung liefern. Fledermäuse sind jedoch zunehmend durch Lebensraumverlust, Krankheiten wie das White-Nose-Syndrom, Klimawandel und erzwungene Interaktionen mit Menschen, bspw. durch die Nähe zu Windenergiestandorten, bedroht. Ihr Schutz ist nicht nur für die Artenvielfalt wichtig, sondern auch für die Stabilität vieler ökologischer Prozesse.

---

Translation of the exemplary version of a provisioned background text for the late stage

---

**Bats – Masters of the night**

Bats are the only mammals capable of active flight. Globally, there are over 1,400 species, playing a crucial role in various ecosystems. Depending on the species, they feed on insects, fruits, nectar, or even blood. Insectivorous bats are valuable pest controllers, as they consume large quantities of insects. Their navigation system, echolocation, allows them to move safely even in complete darkness. They emit high-frequency sound waves, which are reflected by objects and provide an acoustic image of their surroundings.

---

However, bats are increasingly threatened by habitat loss, diseases such as White-Nose Syndrome, climate change, and forced interactions with humans, for instance, due to proximity to wind energy sites. Protecting bats is not only vital for biodiversity but also for maintaining the stability of many ecological processes.

Example research questions:

- How does echolocation influence the hunting strategies of different bat species in various habitats?
- What role do bats play as pest controllers in agricultural regions?
- How do nectar-feeding bats interact with pollinator plants, and what are the implications for biodiversity?
- How effective is the application of antimycotics in bat caves in curbing the spread of White-Nose Syndrome?
- How does climate change impact the migration patterns and distribution of bat populations?
- How does local bus-stop advertisement change the public awareness about bat conservation among students?
- Do ultra-sound generators installed over a period of 6 months help to reduce deaths of bat from wind energy facilities?